\documentclass[letterpaper, 10 pt, conference]{ieeeconf}  

\IEEEoverridecommandlockouts                              

\usepackage{graphics}
\usepackage{epsfig} 
\usepackage{amsmath}
\usepackage{subfigure}          
\usepackage{amssymb}
\usepackage{breqn}
\graphicspath{{./figures/}}
\usepackage{caption}
\usepackage{float}
\usepackage{psfrag}

\title{\LARGE \bf
Control of Large Swarms via Random Finite Set Theory*
}

\author{Bryce Doerr$^{1}$ and Richard Linares$^{2}$
\thanks{*This work was supported by the JPL STTR Program.}
\thanks{$^{1}$Bryce Doerr, Graduate Student, Department of Aerospace Engineering and Mechanics, University of Minnesota, Minneapolis, MN 55455, USA {\tt\small doerr024@umn.edu}}%
\thanks{$^{2}$Richard Linares, Assistant Professor, Department of Aerospace Engineering and Mechanics, University of Minnesota, Minneapolis, MN 55455, USA {\tt\small rlinares@umn.edu}}%
}

\begin{document}

\maketitle
\thispagestyle{empty}
\pagestyle{empty}

\begin{abstract}


Controlling large swarms of robotic agents has many challenges including, but not limited to, computational complexity due to the number of agents, uncertainty in the functionality of each agent in the swarm, and uncertainty in the swarm's configuration. This work generalizes the swarm state using Random Finite Set (RFS) theory and solves the control problem using model predictive control which naturally handles the challenges. This work uses information divergence to define the distance between swarm RFS and a desired distribution. A stochastic optimal control problem is formulated using a modified $L_2^2$ distance. Simulation results show that swarm densities converge to a target destination, and the RFS control formulation can vary in the number of target destinations.


\end{abstract}

\section{INTRODUCTION}

Control of large robotic groups or swarms is currently an area of great interest in controls research. These robotic groups are usually tiny robots with limited actuators that do specific tasks in the formation. For example, the group can use its combined effort to grasp or move in the environment \cite{1}. Specifically, in NASA applications, swarm control of space systems including satellites and rovers can be used for exploration of asteroids and other celestial bodies of interest \cite{2}. In military applications, swarms of UAVs can be used for border patrol, search and rescue, surveillance, communications relaying, and mapping of hostile territory \cite{3}. 
In reviewing these applications, swarm control of large groups of swarming agents are required. 

Changing how the swarm is modeled can alleviate computational complexity for different control methods. Previously, the swarm/potential model using the random finite set (RFS) formalism was used to describe the temporal evolution of the probabilistic description of the robot swarm to promote decentralized coordination \cite{28}. By using a measure-value recursion of the RFS formalism for the agents of the swarm, the dynamics of the swarm in time can be determined with computational efficiency.

Several control techniques have been implemented on swarms previously. In centralized control, one agent in the swarm is the designated controller and manages the execution of other agents in the swarm \cite{5}. Thus, it oversees the other agents' system processes. This type of control is the easiest to implement on a swarm of robots, and it solves the problem of controlling the swarm to a desired state. Unfortunately, this type of control suffers from two main problems. As the number of agents of the swarm increases, the computational workload becomes more difficult. This is especially difficult when the swarm robots are low cost robotic agents. Another issue that arises is that centralized control is not robust against failures of other agents within the swarm \cite{6}. Since swarm robots on a level of a thousand agents are low cost units, communication, actuation, and sensing are performed with less reliability. Thus, centralized control may not be a viable option for these systems.  


Probabilistic swarm guidance has also been used to enable the swarm to converge to target distributions \cite{9}. Probabilistic swarm guidance controls a swarm density distribution through distributed control, so that each agent determines its own trajectory while the swarm converges to the target distribution. Distributed control is defined as reformulating a control problem as a set of interdependent subproblems and solving these subproblems \cite{8}. Probabilistic swarm guidance solves issues that involve controlling an increasing number of individual agents which is computationally complex by controlling the swarm density distribution of the agents. Using optimal transport, convergence is achieved faster than a homogeneous Markov chain approach and the cost function is minimized \cite{9,20,21}. Another method is the use of the inhomogeneous Markov chain \cite{16,17}. In this approach, the agents move in a decentralized fashion which is like the homogeneous Markov chain, but the algorithm allows communication with other agents to reach and settle at the target destination. These algorithms are robust to external disturbances since the algorithms themselves use the swarm density distribution. 

Velocity field generation for swarm control is a non-optimal decentralized control method for swarms that synthesizes smooth velocity fields as a function of time and position \cite{29}. With a designated target distribution, the swarm follows the velocity field using the heat equation to convergence by using local agent position information to estimate its local density. The advantage of this method is that the agents facilitate collision avoidance and move in a smoother manner than previous Markov chain approaches \cite{20}. Unfortunately, the use of the heat equation diffuses the agents in a locally uniform manner to the target density. Therefore, this is a non-optimal method of controlling the swarm to a target distribution.

Model predictive control (MPC) has been heavily studied for nonlinear systems and
for applications including spacecraft maneuvering and attitude control \cite{22,25,26}. Decentralized MPC has also been studied for thousands of low cost spacecraft with limited capabilities \cite{10}. MPC computes the control input by solving a finite horizon problem. This allows it to optimize the current time step while taking the future time steps into account. 


One last decentralized approach to controlling swarms is using sequential convex programming \cite{18}. Sequential convex programming uses multiple iterations to maintain accuracy of the convex approximations of the constraints which create more efficient trajectories. Uncertainties in the trajectories are accounted for when the algorithm is tuned. Thus, this provides the robustness for the swarm while pushing the agents to the designated targets while using this algorithm in combination with MPC in real time. Morgan also used sequential convex programming to do target assignment (mapping of agents to targets) and trajectory generation for varying swarm sizes through time \cite{15}. This method is viable for swarms, but through RFS control, target assignment is not necessary.


\section{Random Finite Set Control Problem Formulation}

The control problem for swarming agents is set up by using RFS theory \cite{19,27}. This theory addresses the decentralized formulation for each agent in the formation. Each agent has the challenge of estimating its local formation configuration and designing a decentralized control policy to achieve some local configuration. It is assumed that each agent within the swarm is identical and using unique identifiers on each agent is unnecessary. Using RFS theory, the number of agents and their states is determined from measurements. The number of agents in the field is denoted by $N_{\text{total}}(t)$ and may be varying as agents enter and leave the field. $X_t= \left\{{\bf x}_{1,t},{\bf x}_{2,t},\cdots,{\bf x}_{N_{total(t)},t} \right\}$ denotes a realization of the RFS distribution for agents. 
At any time, $t$, the RFS probability density function can be written as
\begin{dmath}\label{rfsstate}
p(X=\left\{{\bf x}_1,{\bf x}_2,\cdots,{\bf x}_n\right\})=p(|X|=n)p(\left\{{\bf x}_1,{\bf x}_2,\cdots,{\bf x}_n\right\}{\mid \lvert X\rvert =n)}.
\end{dmath}
The control sequence is also defined by a RFS in the form $U_t= \left\{{\bf u}_{1,t},{\bf u}_{2,t},\cdots,{\bf u}_{N_{total(t)},t} \right\}$ and a RFS probability density given by
\begin{dmath}\label{rfsstate}
p(U=\left\{{\bf u}_1,{\bf u}_2,\cdots,{\bf u}_n\right\})=p(|U|=n)p(\left\{{\bf u}_1,{\bf u}_2,\cdots,{\bf u}_n\right\}{\mid \lvert U\rvert =n)},
\end{dmath}
since the number of the agents in the field to be controlled is also varying.
The Bayesian multi-target state estimation problem generally has no closed-form solution, but when the form $p(|X|=n)$ and $p(\left\{{\bf x}_1,{\bf x}_2,...,{\bf x}_n\right\}{\mid \lvert X\rvert =n)}$ are assumed, approximate solutions can be determined by the Probability Hypothesis Density (PHD) filter. Using the PHD filter assumption, the RFS is represented by its first moment (intensity function $\nu({\bf x},t)$) \cite{19}. $\nu({\bf x},t)$ represents the probability of finding an agent in a region of state space. The PHD is defined over the single agent's state space and has the property that the expected number of agents over a region is simply the integral of the PHD over that region. Propagation of the PHD can be determined if the agents are assumed to be independent and identically distributed with the cardinality of the agent set that is Poisson distributed \cite{27}. It is noted that the assumptions made by the PHD filter are strong assumption for swarming robotics. However, this is a good starting point for initial proof-of-concept study. 
The PHD recursion for a general intensity function, $v_t({\bf x})$, is given by

\begin{equation}\label{rfs1}
\bar{v}_t({\bf x})=b({\bf x})+\int p_s(\zeta)f({\bf x}|\zeta)v(\zeta)d\zeta,
\end{equation}
where $b({\bf x})$, $p_s(\zeta)$, and $f({\bf x}|\zeta)$ are the agents' birth, survival probability, and target motion models respectively and $\zeta$ is the previous state \cite{27}. The bar on $\bar{v}_t({\bf x})$ denotes that the PHD is conditioned on measurements $Z_{t-1}$. For the time update, the equation is given by

\begin{dmath} \label{rfs2}
{v_t({\bf x})=(1-p_d({\bf x}))\bar{v}_t({\bf x})}+\sum_{z\in Z_t}\frac{p_d({\bf x})g(z_t|{\bf x})\bar{v}_t({\bf x})}{c(z)+\int p_d(\zeta)g(z_t|\zeta)\bar{v}_t(\zeta)d\zeta},
\end{dmath}
where $p_d({\bf x})$, $g(z_t|{\bf x})$, and $c(z)$ are the probability of detection, likelihood function, and clutter model of the sensor respectively \cite{27}. By using this recursion, the swarm probabilistic description can be updated. Unfortunately, Eqs.~\eqref{rfs1} and \eqref{rfs2} do not contain a closed-form solution and the numerical integration suffers from higher computational time as the states increase. Fortunately, a closed-form solution exists if it is assumed that the survival and detection probabilities are state independent, and the intensities of the birth and spawn RFSs are Gaussian mixtures of the form
\begin{equation}\label{funcf}
\nu({\bf x},t)=f({\bf x})=\sum_{i=1}^{N_f}w_f^{(i)}\mathcal{N}\left({\bf x};{\bf m}_f^i,P_f^i\right),
\end{equation}
\begin{equation}\label{funcg}
\bar{\nu}({\bf x},t)=g({\bf x})=\sum_{i=1}^{N_g}w_g^{(i)}\mathcal{N}\left({{\bf x};\bf m}_g^i,P_g^i\right),
\end{equation}
where $w$ are the weights and $\mathcal{N}\left({\bf x};{\bf m}^i,P^i\right)$ is the probability density function of a multivariate Gaussian distribution with a mean and covariance respectively. Note that closed form solutions using Gaussian mixtures exist for cases without the state independent assumption. Additionally, $\sum_{i=1}^{N_f}w_f^{(i)}=N_{\text{total}}(t)$ and $\sum_{i=1}^{N_g}w_g^{(i)}=\bar{N}_{\text{total}}(t)$ where $\bar{N}_{\text{total}}(t)$ is the desired number of agents. The intensity function $\nu({\bf x},t)$ is in terms of the swarm state while $\bar{\nu}({\bf x},t)$ is in terms of the desired state.
The swarm intensity function can be propagated through updates on the mean and covariance as given by
\begin{equation}
{\bf m}_{f,k+1}^i=A_k{\bf m}_{f,k}^i+B_k{\bf u}_{f,k}^i,
\end{equation}
\begin{equation}
P_{f,k+1}^i=A_kP_{f,k}^iA_k^T+Q_k,
\end{equation}
where $Q_k$ is process noise. Then given the Gaussian mixture intensities assumption, a control variable is calculated for each component ${\bf u}_{f,k}^i$. Additionally, each Gaussian mixture component can represent many agents since the intensity function integrates to the total number of agents. Note that although linear dynamics is used, the RFS-based cost function is nonquadratic.

Each individual swarm agent will run a local PHD observer to estimate the state of the swarm by modeling the swarm as a distribution. Thus, using RFS theory, it is assumed that the individual swarm agents form an intensity function that is a Gaussian mixture density in which the means and covariances of the Gaussian mixture are propagated and controlled. An optimal control problem is set up that tracks a desired swarm formation by minimizing its control effort in the following objective function 
\begin{dmath}\label{distance}
J({\bf u})=\int_{0}^TE\left\{{\bf u}^{\intercal}R{\bf u}\right\}+D(\nu({\bf x},t),\bar{\nu}({\bf x},t))dt,
\end{dmath}
where $\bar{\nu}({\bf x},t)$ is the desired formation and ${\bf u}$ is the control effort for the Gaussian mixture swarm. Both $\nu({\bf x},t)$ and $\bar{\nu}({\bf x},t)$ are defined over the complete state space which include position and velocity parameters. $D(\nu({\bf x},t),\bar{\nu}({\bf x},t))$ is the distance between these Gaussian mixtures which has several closed-form solutions.


\section{Distributional Distance Based-Cost}
\subsection{Cauchy-Schwarz Divergence}
The Cauchy-Schwarz divergence is based on the Cauchy-Schwarz inequality for inner products, and it is defined for two random vectors with probability densities $f$ and $g$ given by 
\begin{equation}\label{cauchy}
D_{CS}(f,g)=\ln\left(\frac{\left\langle f,g\right\rangle}{\|f\|\|g\|}\right),
\end{equation}
where $\left\langle \cdot,\cdot \right\rangle$ is the $L_2^2$ inner product over the densities. The argument of the logarithm is non-negative since probability densities are non-negative, and it does not exceed one by the Cauchy-Schwarz inequality. The Cauchy-Schwarz divergence can be interpreted as an approximation to the Kullback-Leibler divergence but has a closed-form expression for Gaussian mixtures \cite{12}. This is useful for calculating the distance between two-point processes represented by intensity functions. The Cauchy-Schwarz divergence between two Poisson point processes with Gaussian mixture intensities is shown in Eq. \eqref{cauchy} when Eqs.~\eqref{funcf} and \eqref{funcg} are substituted inside.
Note that in the control formulation used, only $\nu({\bf x},t)$ is assumed to depend on the control ${\bf u}$. Therefore, the term that depends only on $\bar{\nu}({\bf x},t)$ is omitted from the objective function since $\bar{\nu}({\bf x},t)$ does not depend on ${\bf u}$. 

\begin{figure}[h]
\begin{centering}
    \subfigure[Cauchy-Schwarz Divergence]{
      \psfrag{Position Shape}[][]{\small{}}
      \psfrag{x}[][]{\small{$x$}}
      \psfrag{y}[][]{\small{$y$}}
      {\includegraphics[width=.2\textwidth]{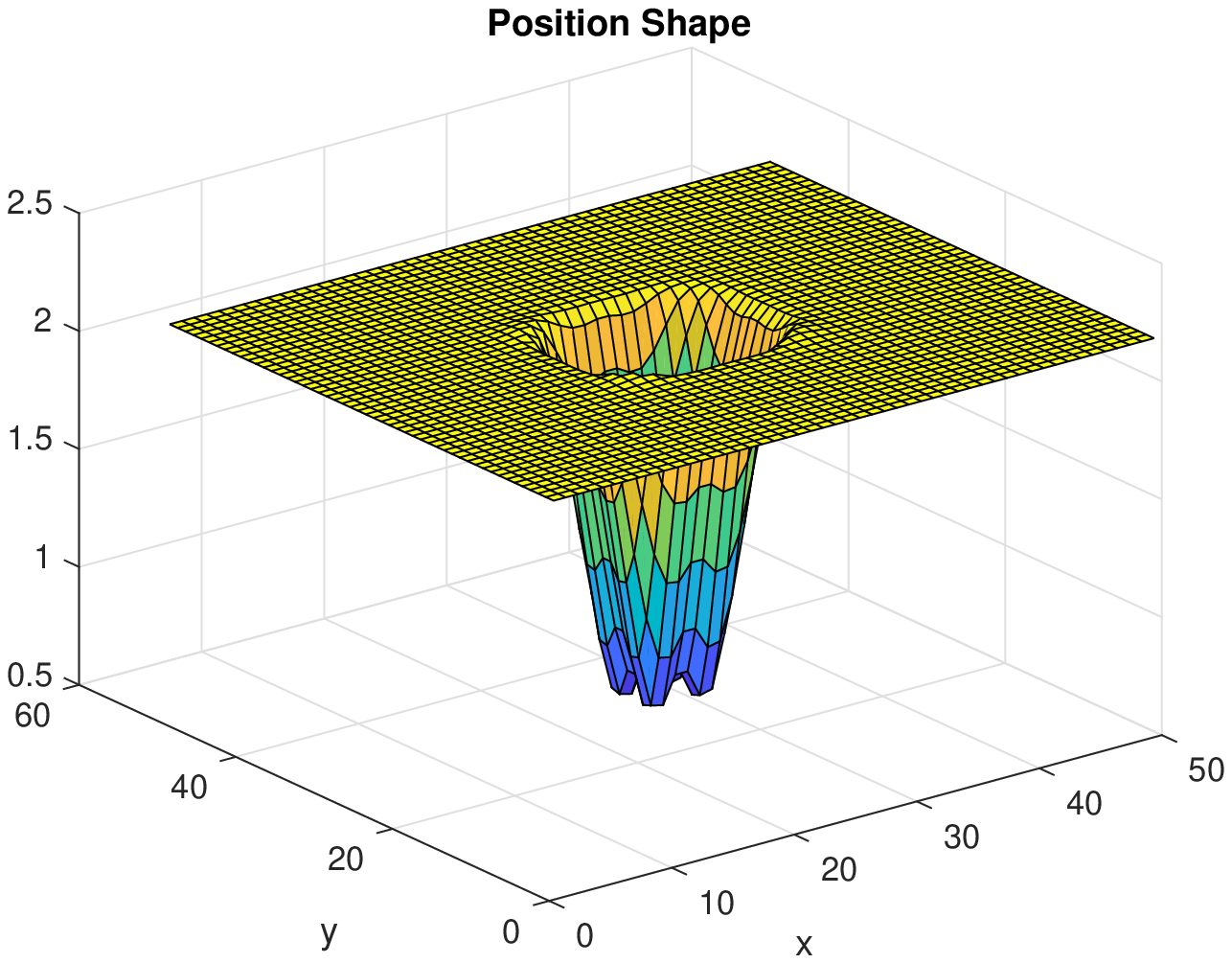}}\label{fig:CS}}
          \subfigure[$L_2^2$ Distance]{
      \psfrag{Position Shape}[][]{\small{}}
      \psfrag{x}[][]{\small{$x$}}
      \psfrag{y}[][]{\small{$y$}}
    {\includegraphics[width=.2\textwidth]{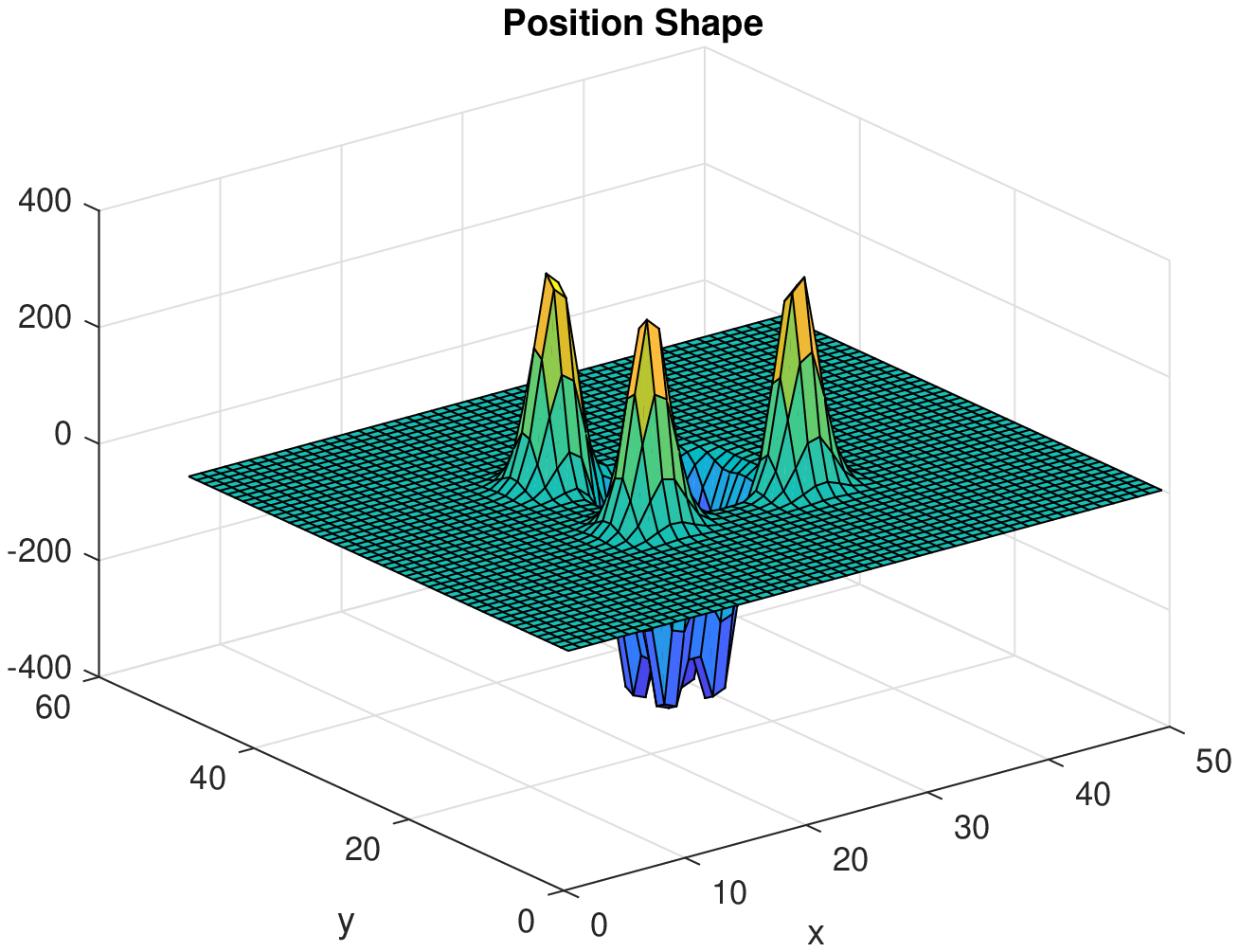}}\label{fig:L2}}  
              \subfigure[$L_2^2$ + Quadratic Term]{
      \psfrag{Position Shape}[][]{\small{}}
      \psfrag{x}[][]{\small{$x$}}
      \psfrag{y}[][]{\small{$y$}}
    { \includegraphics[width=.2\textwidth]{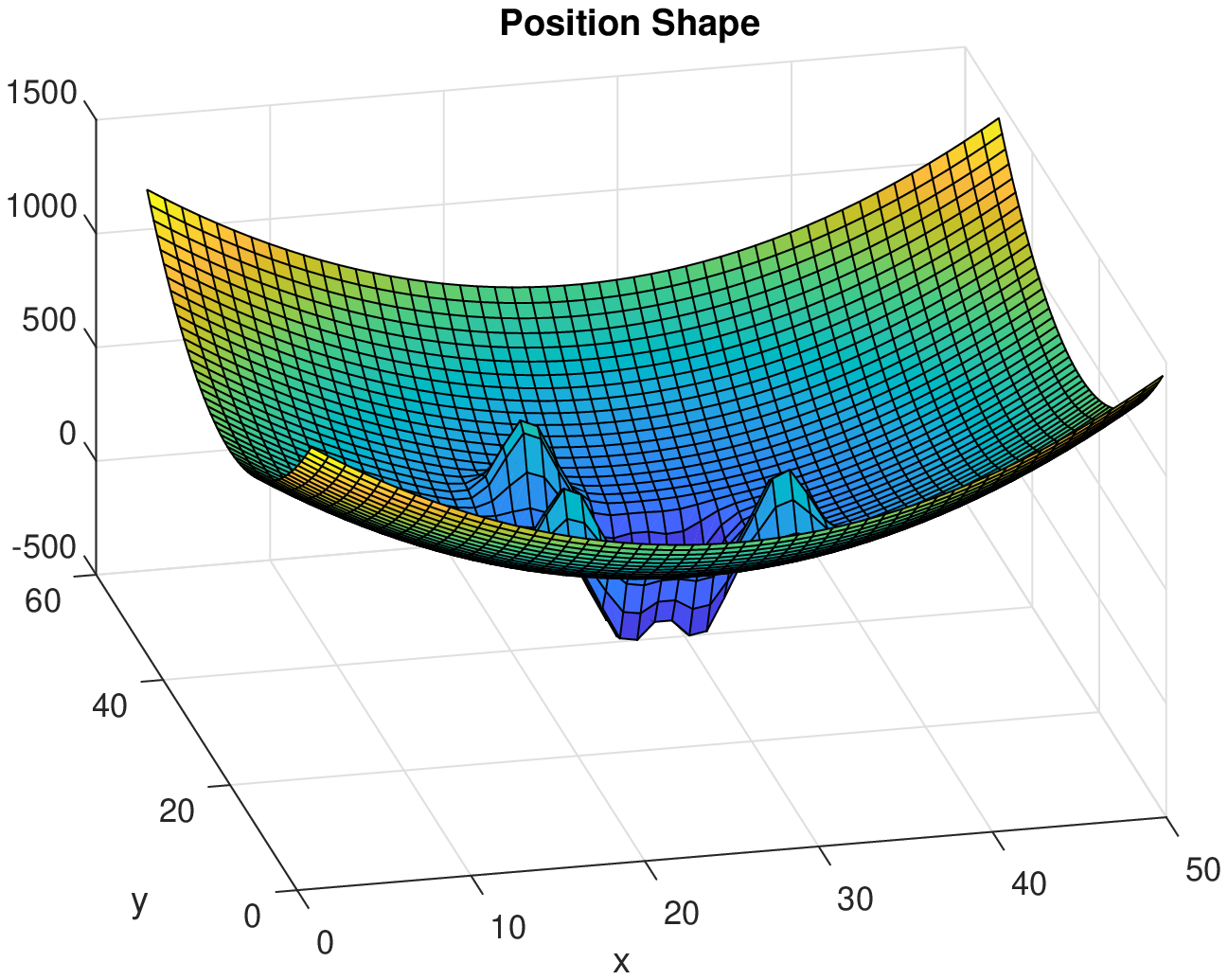}}\label{fig:L2q}} 
        \caption{Surface Plots of the RFS divergences.}\label{contours}
\end{centering}
\end{figure}

Figure \ref{fig:CS} shows the surface plot using the Cauchy-Schwarz divergence for 4 densities in the swarm at an initial time instance. Each initial state has hills while the destination states have valleys. Thus, each density will repel each other while moving towards the final state through time. If the initial state is too large compared to the destination states, it will take longer for the 4 densities to converge to the destination values. Also, the repelling effect due to the hills are relatively small. Thus, the Cauchy-Schwarz divergence may not be the fastest converging solution for the objective function minimization.
\vspace{-1.95ex}
\vspace{-.25ex}
\subsection{$L_2^2$ Distance}
Alternatively, the distance between two Poisson point processes with Gaussian mixture intensities can be determined by using the $L_2^2$ distance between the probability densities. The $L_2^2$ is given by
\begin{equation}
D_{L_2^2}(f,g)=\int\left( f-g\right)^2d{\bf x}.
\end{equation}
Figure \ref{fig:L2} shows surface plot using the $L_2^2$ distance for a 4-density swarm at an initial time instance. Each initial state has more defined hills compared to the Cauchy-Schwarz divergence. Thus, the initial states have a stronger repel effect from each other. Also, the destination states have large valleys that create a large attraction effect for each initial state to move to. Thus, the optimization solution will be faster in the $L_2^2$ distance case. Unfortunately, the $L_2^2$ distance suffers from a similar issue to the Cauchy-Schwarz divergence. If the initial conditions increase farther away from the destination states, the optimization may take much longer due to a flat surface away from the destination states.

\subsection{$L_2^2$ Distance with Quadratic Term}
The issue of convergence remains for the $L_2^2$ distance when the initial states are far larger than the destination states. To achieve faster convergence, an additional term is added to the $L_2^2$ distance to shape the gradient descent through a quadratic term as given by 
\begin{dmath}\label{modifiedL2}
{D_{L_2^2mod}(f,g)=D_{L_2^2}(f,g)}-{ \sum_{j=1}^{N_g}\sum_{i=1}^{N_f}w_g^{(j)}w_f^{(i)}\ln\left(\mathcal{N}({\bf m}_g^j;{\bf m}_f^i,P_g^i+P_f^j)\right)}.
\end{dmath}
Note that this term is referred as quadratic, although it may be more appropriate to call it quadratic-like.
Figure \ref{fig:L2q} shows the surface plot using the modified $L_2^2$ distance for a 4-density swarm in Eq. \eqref{modifiedL2}. Compared to the $L_2^2$ distance, the initial and destination states provide the hills and valleys necessary to obtain convergence. However, as the initial states move outwards, the surface map decreases in a quadratic fashion instead of staying flat. Thus, the optimization can occur at any point to reach convergence in a timely manner.

To solve the minimization problem for the objective function, traditional LQR based solutions are not applicable since the terms inside the objective functions are nonquadratic \cite{11}. 
The minimization of the objective function in discrete time is
\begin{dmath}\label{overalleqn}
{\min_{{\bf u}_k,k=1,...,T}J({\bf u})=\sum_{k=1}^T\ E\{{\bf u}_k^{\intercal}R{\bf u}_k\}}+\sum_{j=1}^{N_f}\sum_{i=1}^{N_f}w_{f,k}^{(j)}w_{f,k}^{(i)}\mathcal{N}({\bf m}_{f,k}^j;{\bf m}_{f,k}^i,P_{f,k}^i+P_{f,k}^j)+\sum_{j=1}^{N_g}\sum_{i=1}^{N_g}w_{g,k}^{(j)}w_{g,k}^{(i)}\mathcal{N}({\bf m}_{g,k}^j;{\bf m}_{g,k}^i,P_{g,k}^i+P_{g,k}^j)-2\sum_{j=1}^{N_g}\sum_{i=1}^{N_f}w_{g,k}^{(j)}w_{f,k}^{(i)}\mathcal{N}({\bf m}_{g,k}^j;{\bf m}_{f,k}^i,P_{g,k}^i+P_{f,k}^j)- \sum_{j=1}^{N_g}\sum_{i=1}^{N_f}w_{g,k}^{(j)}w_{f,k}^{(i)}\ln\left(\mathcal{N}({\bf m}_{g,k}^j;{\bf m}_{f,k}^i,P_{g,k}^i+P_{f,k}^j)\right),
\end{dmath}
\begin{equation}\label{qwer}
\text{Subject} \hspace{5pt}\text{to}:{\bf m}_{f,k+1}^i=A_k{\bf m}_{f,k}^i+B_k{\bf u}_{f,k}^i,
\end{equation}
where ${\bf u}_k=[({\bf u}_{f,k}^1)^T,\cdots,({\bf u}_{f,k}^{N_f})^\intercal]^\intercal$ is the collection of all control variables. 

\section{Receding Horizon Dynamic Programming}
An optimal solution based on minimizing the objective uses MPC or receding horizon control. This uses dynamic programming techniques to obtain an optimal control solution ${\bf u}$ from the objective function and has been implemented on swarm control previously \cite{15}. 

Conceptually, at a time t, the knowledge of the system model is used to derive a sequence ${\bf u}(t|t), {\bf u}(t+1|t), {\bf u}(t+2|t),\cdots, {\bf u}(t+T_p|t)$ where $T_p$ is the finite horizon from the current state $\bar{\bf x}(k)$ \cite{14}. With the input sequence, the state is moved forward in time by the control horizon, $T_c$; usually one time-step. Then the same strategy is repeated for time $t+1$. $T_p$ can be chosen to be either small or large. As $T_p$ increases, the degrees of freedom in the optimization increases which can slow down the algorithm considerably even though more of the future reference trajectory would be useful to bring the output closer to the reference. With a smaller $T_p$, the computation time will be faster, but the optimization may be more suboptimal. Thus, the swarm may not converge to the desired configuration. 

For the RFS control formulation, a ${\bf u}$ that controls the swarm densities through their statistics (mean and covariance) was found by minimizing the objective as given by Eq. \eqref{overalleqn} and \eqref{qwer}. This was done using MATLAB's fminunc solver and the Quasi-Newton algorithm \cite{30}. Then, the agents in the swarm were initialized to the closest density using the Mahalanobis distance given by
\begin{equation}
D_M({\bf x},{\bf m}_{f,k}^i)=\sqrt{({\bf x}-{\bf m}_{f,k}^i)^{\intercal}P^{-1}({\bf x}-{\bf m}_{f,k}^i)},
\end{equation}
which measures the distance of the agents to the means of the densities. As the swarm evolved, this distance determined which units belong to a given component. Agents were controlled according to their placement in each density through an open loop method using the MPC control input obtained for each density. Although an open loop method was used, feedback control can be used if the state estimates are determined from the PHD filter.

The great advantage of using MPC is that the algorithm can handle nonlinearities in the objective function. Unfortunately, numerically solving an optimization at each time-step is not particularly fast, so it can be difficult to run optimization problems in real-time to control swarm robots towards a target state.

%


\section{Results}
In the simulation experiment, a 4-density swarm on a 2-D plane was initialized in a square grid where the densities 1, 2, 3, and 4 were defined counterclockwise starting on the first quadrant. With the 4-density swarm, four different test cases were implemented to bring the densities to the target trajectories and to test the control theory involved from the control formulation. The first test case shows the limitations of using only the $L_2^2$ distance with varying initial conditions in a square grid and four target destinations located at ($\pm$1,$\pm$1). The last three cases used the $L_2^2$ distance with a quadratic term and varying target destinations. In Case 2, four target destinations are located at ($\pm$1,$\pm$1). For Case 3, three target destinations are located at ($\pm$1,1) and (-1,-1). Lastly in Case 4, five target destinations are located at ($\pm$1,$\pm$1) and (0,0). The receding horizon dynamic programming was used to determine the control effort and trajectories for the simulation, but other methods including iterative LQR can be used too \cite{11}.
\begin{figure}[!ht]
\centering
\psfrag{xPos. 1}[][]{\scriptsize{${\bf m}_{1,x}$}}
\psfrag{yPos. 1}[][]{\scriptsize{${\bf m}_{1,y}$}}
\psfrag{xPos. 2}[][]{\scriptsize{${\bf m}_{2,x}$}}
\psfrag{yPos. 2}[][]{\scriptsize{${\bf m}_{2,y}$}}
\psfrag{xPos. 3}[][]{\scriptsize{${\bf m}_{3,x}$}}
\psfrag{yPos. 3}[][]{\scriptsize{${\bf m}_{3,y}$}}
\psfrag{xPos. 4}[][]{\scriptsize{${\bf m}_{4,x}$}}
\psfrag{yPos. 4}[][]{\scriptsize{${\bf m}_{4,y}$}}
\psfrag{time (s)}[][]{\scriptsize{Time (s)}}
\includegraphics[keepaspectratio, width=0.45\textwidth]{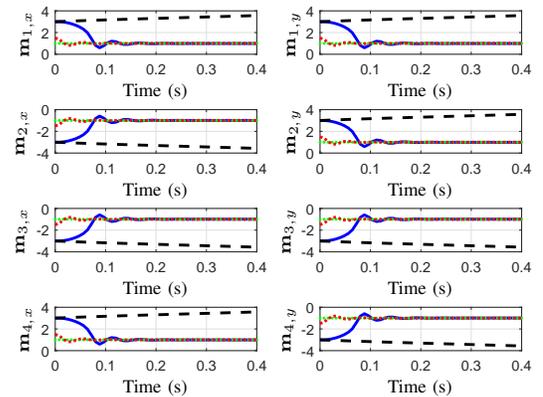}
\caption{The position responses of density statistics (means) are shown for Case 1 and Case 2. Black-dashed and red-dotted lines are the responses for $L_2^2$ distance with ($\pm$3,$\pm$3)) and ($\pm$1.5,$\pm$1.5)) initial conditions respectively. The blue-solid lines are the responses for the $L_2^2$+Quadratic term with initial conditions of ($\pm$3,$\pm$3)). The green-dotted lines are the target destinations for both cases.}\label{fig:Case01}
\centering
\end{figure}
\subsection{Case 1: $L_2^2$}
\begin{figure}[!ht]
\centering
\psfrag{x}[][]{\scriptsize{$x$}}
\psfrag{y}[][]{\scriptsize{$y$}}
\psfrag{time = 0.00 s}[][]{\scriptsize{time = 0.00 s}}
\psfrag{(a) Initial conditions of (+3,+3)}[][]{\tiny{(a) Initial conditions of ($\pm$3,$\pm$3)}}
\psfrag{(b) Initial conditions of (+1.5,+1.5)}[][]{\tiny{(b) Initial conditions of ($\pm$1.5,$\pm$1.5)}}
\psfrag{time = 0.10 s}[][]{\scriptsize{time = 0.10 s}}
\psfrag{time = 0.40 s}[][]{\scriptsize{time = 0.40 s}}
\includegraphics[keepaspectratio, width=0.41\textwidth]{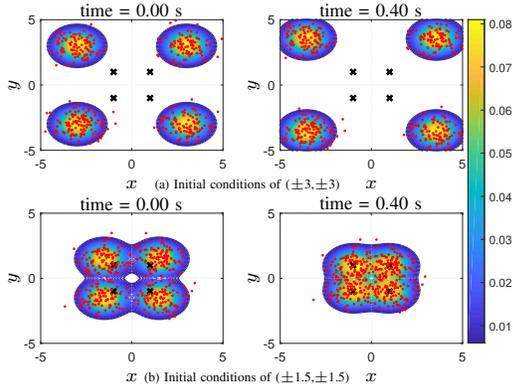}
\caption{Case 1: Trajectories of the swarm with individual agents (red dots) and four target destinations (black x's). 
}\label{Case0}
\centering
\end{figure}

For Case 1, four swarm densities are controlled to move towards the target destinations at initial conditions farther away (square grid at ($\pm$3,$\pm$3)) and closer to (square grid at ($\pm$1.5,$\pm$1.5)) the target destinations as shown by mean responses given by the black-dashed and red-dotted lines in Figure \ref{fig:Case01} respectively. From the trajectory snapshots given by Figure \ref{Case0}(a), initial conditions that are far from the target destinations do not have a converging control solution. From the surface visualization in Figure \ref{fig:L2}, the general plane is flat in areas away from the target destinations and states of the density. Therefore, optimization using MPC is more difficult in these flat areas and may not converge to a solution. If the densities are initialized much closer to the target destinations as shown in Figure \ref{Case0}(b), the flatness in the general plane is minimal, and the optimization step in MPC converges to a solution. By using the $L_2^2$ distance, converging control solutions can only be found for initial conditions and target destinations that are close.

\subsection{Case 2: $L_2^2$+Quadratic Term, Four Target Destinations}
In Case 2, four swarm densities move towards the four target destinations given by the blue-solid lines (mean responses) in Figure \ref{fig:Case01}. 
Figure \ref{fig:Case1sub2} shows the trajectory snapshots and final states of each of the swarm densities during the simulation. The target destinations are plotted as black x's. The red dots are the individual swarm agents that form the Gaussian mixture densities. From the figure, all four densities converge to the target destinations in approximately 0.17 seconds and approximately 0.03 of steady-state error between the densities' position to the target destinations. In comparison to Case 1, Figure \ref{fig:Case01} shows that for small distances between the initial state and the target destination, the $L_2^2$ distance is sufficient for state convergence, but as the distance increases, the $L_2^2$ distance diverges away. By adding the quadratic term to $L_2^2$, the optimization step can directly determine the minimum for the control solution shown in Figure \ref{fig:L2q}. Therefore, the target destinations attract the swarm densities at distances that fail for only $L_2^2$ distance given by Figure \ref{fig:Case01}.
\begin{figure}[!ht]
\centering
\psfrag{x}[][]{\scriptsize{$x$}}
\psfrag{y}[][]{\scriptsize{$y$}}
\psfrag{time = 0.00 s}[][]{\scriptsize{time = 0.00 s}}
\psfrag{time = 0.05 s}[][]{\scriptsize{time = 0.05 s}}
\psfrag{time = 0.10 s}[][]{\scriptsize{time = 0.10 s}}
\psfrag{time = 0.40 s}[][]{\scriptsize{time = 0.40 s}}
\includegraphics[keepaspectratio, width=0.40\textwidth]{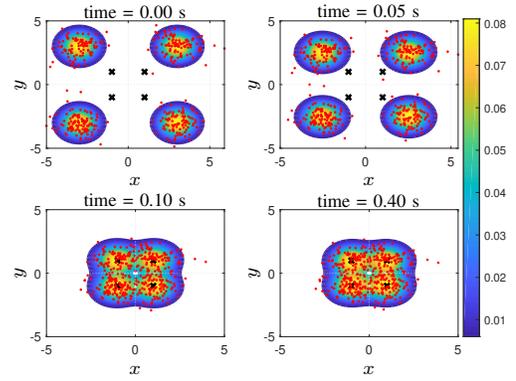}\label{fig:Case1sub2}
\caption{Case 2: Trajectories of the swarm with individual agents (red dots) and four target destinations (black x's). 
}\label{fig:Case1sub2}
\centering
\end{figure}

\subsection{Case 3: $L_2^2$+Quadratic Term, Three Target Destinations}
Case 3 illustrates the effect of three target destinations on the final trajectories of the four swarm densities.
From Figure \ref{fig:Case2sub2}, it can be visually shown where the swarm densities are located relative to the target destinations at each time step. The trajectories for density 1 and density 3 reach their target, but densities 2 and 4 reach the third target with approximately 0.30 of steady-state error. The results obtained follow directly from the RFS control theory using the $L_2^2$ plus quadratic distance term. By using this $L_2^2$ with a quadratic term in the objective function, the individual densities will attract towards the target destinations while repulsing away from each other. This can be shown in the surface map shown in Figure \ref{fig:L2q} where the hills are areas of repulsion and valleys are areas of attraction. Thus, for Case 3, densities 2 and 4 are attracted to the same target, but they stay away from each other which causes both to share the same target at a distance away. 

\vspace{-1.95ex}
\begin{figure}[!ht]
\centering
\psfrag{x}[][]{\scriptsize{$x$}}
\psfrag{y}[][]{\scriptsize{$y$}}
\psfrag{time = 0.00 s}[][]{\scriptsize{time = 0.00 s}}
\psfrag{time = 0.05 s}[][]{\scriptsize{time = 0.05 s}}
\psfrag{time = 0.10 s}[][]{\scriptsize{time = 0.10 s}}
\psfrag{time = 0.40 s}[][]{\scriptsize{time = 0.40 s}}
\includegraphics[keepaspectratio, width=0.40\textwidth]{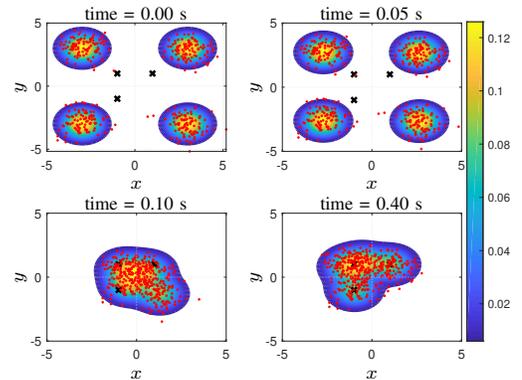}\label{fig:Case2sub2}
\caption{Case 3: Trajectories for the swarm with individual agents (red dots) and three target destinations (black x's).
}\label{fig:Case2sub2}
\centering
\end{figure}
\vspace{-3ex}

\subsection{Case 4: $L_2^2$+Quadratic term, Five Target Destinations}
For Case 4, four swarm densities are attracted to five target destinations.
The trajectory snapshots of the densities are visually shown relative to the five target destinations in Figure \ref{fig:Case3sub2}. All densities have a steady state error of approximately 0.17 which follow the theory as expected. Since the swarm densities are far from each other, the effects of repulsion are minimal. Also, the densities are attracted to the four target destinations that make up a square, but they are also attracted to the target destination at the origin. This is due to the minimization of the objective function that has a $L_2^2$ and a quadratic term where the individual densities will attract towards the target destinations. Since there is an additional target destination of attraction at the origin, all four swarm densities are affected by the origin as they are moving towards the 4 square target destinations. Thus, compared to Case 2 with only 4 target destinations, the densities in this case will have steady state error due to attraction of the target destination that has no density.


\begin{figure}[!ht]
\centering
\psfrag{x}[][]{\scriptsize{$x$}}
\psfrag{y}[][]{\scriptsize{$y$}}
\psfrag{time = 0.00 s}[][]{\scriptsize{time = 0.00 s}}
\psfrag{time = 0.05 s}[][]{\scriptsize{time = 0.05 s}}
\psfrag{time = 0.10 s}[][]{\scriptsize{time = 0.10 s}}
\psfrag{time = 0.40 s}[][]{\scriptsize{time = 0.40 s}}
\includegraphics[keepaspectratio, width=0.45\textwidth]{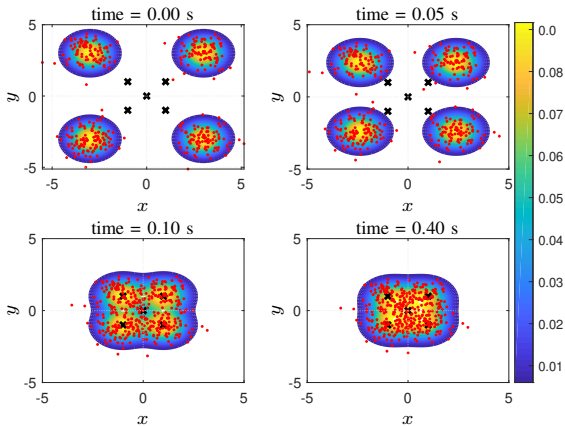}\label{fig:Case3sub2}
\caption{Case 4: Trajectories for the swarm with individual agents (red dots) and five target destinations (black x's).
}\label{fig:Case3sub2}
\centering
\end{figure}
\vspace{-3ex}
\section{Conclusions}
The objective of this paper was to formulate a method to control swarming formations by MPC and RFS theory. By setting up the problem using information divergence to define the distance between the swarm RFS and a desired distribution, an optimal control problem was found that tracked a linear system with a nonquadratic objective function using MPC that approximates the optimal control solution that minimizes the objective function. 

\vspace{-1.5ex}
\section*{ACKNOWLEDGMENT}
The authors wish to acknowledge support by the National Aeronautics and Space Administration under Contract Number NNX16CP45P issued through the NASA STTR Program by the Jet Propulsion Laboratory (JPL) and led by Amir Rahmani at JPL. The authors wish to acknowledge useful conversations related to satellite technologies with Chuck Hisamoto, Vaughn Weirens, and Suneel Sheikh of ASTER Labs, Inc.


\vspace{-3ex}
\begin{thebibliography}{5}
\bibitem{1}
Kube, C. Ronald, and Hong Zhang. ``Collective robotics: From social insects to robots." Adaptive behavior 2.2 (1993): 189-218.
\bibitem{2}
Vassev, Emil, Mike Hinchey, and Joey Paquet. ``Towards an ASSL specification model for NASA swarm-based exploration missions." Proceedings of the 2008 ACM symposium on Applied computing. ACM, 2008. 
\bibitem{3}
Ryan, Allison, et al. ``An overview of emerging results in cooperative UAV control." Decision and Control, 2004. CDC. 43rd IEEE Conference on. Vol. 1. IEEE, 2004.
\bibitem{4}
Rubenstein, Michael, Alejandro Cornejo, and Radhika Nagpal. ``Programmable self-assembly in a thousand-robot swarm." Science 345.6198 (2014): 795-799.
\bibitem{8}
Lunze, Jan. Feedback control of large-scale systems. London: Prentice-Hall, (1992): 9-11.
\bibitem{28}
Pace, M., M. Birattari, and M. Dorigo. ``The swarm/potential model: Modeling robotics swarms with measure-valued recursions associated to random finite sets." IEEE Transactions on Robotics, page submitted (2013).
\bibitem{5}
Centralized Control. Ian Sommerville , 2008. Web. 05 May 2017.
\bibitem{6}
Mondada, Francesco, et al. ``The cooperation of swarm-bots: Physical interactions in collective robotics." IEEE Robotics \& Automation Magazine 12.2 (2005): 21-28.
\bibitem{9}
Bandyopadhyay, Saptarshi, Soon-Jo Chung, and Fred Y. Hadaegh. ``Probabilistic swarm guidance using optimal transport." Control Applications (CCA), 2014 IEEE Conference on. IEEE, 2014.
\bibitem{20}
Acikmese, Behçet, and David S. Bayard. ``A markov chain approach to probabilistic swarm guidance." American Control Conference (ACC), 2012. IEEE, 2012.
\bibitem{21}
Chattopadhyay, Ishanu, and Asok Ray. ``Supervised self-organization of homogeneous swarms using ergodic projections of markov chains." IEEE Transactions on Systems, Man, and Cybernetics, Part B (Cybernetics) 39.6 (2009): 1505-1515.
\bibitem{16}
Bandyopadhyay, Saptarshi, Soon-Jo Chung, and Fred Y. Hadaegh. ``Inhomogeneous Markov chain approach to probabilistic swarm guidance algorithm." 5th Int. Conf. Spacecraft Formation Flying Missions and Technologies. 2013.
\bibitem{17}
Hadaegh, Fred Y., Soon-Jo Chung, and Harish M. Manohara. ``On development of 100-gram-class spacecraft for swarm applications." IEEE Systems Journal 10.2 (2016): 673-684.
\bibitem{29}
Eren, Utku, and Behçet A\c{c}{\i}kme\c{s}e. ``Velocity field generation for density control of swarms using heat equation and smoothing kernels." IFAC-PapersOnLine 50.1 (2017): 9405-9411.
\bibitem{22}
Camacho, Eduardo F., and Carlos Bordons Alba. Model predictive control. Springer Science \& Business Media, 2013.
\bibitem{25}
Di Cairano, S., H. Park, and I. Kolmanovsky. ``Model predictive control approach for guidance of spacecraft rendezvous and proximity maneuvering." International Journal of Robust and Nonlinear Control 22.12 (2012): 1398-1427.
\bibitem{26}
Manikonda, V., et al. ``A model predictive control-based approach for spacecraft formation keeping and attitude control." American Control Conference, 1999. Proceedings of the 1999. Vol. 6. IEEE, 1999.
\bibitem{10}
Morgan, Daniel, Soon-Jo Chung, and Fred Y. Hadaegh. ``Decentralized model predictive control of swarms of spacecraft using sequential convex programming." Advances in the Astronautical Sciences 148 (2013): 1-20.
\bibitem{18}
Morgan, Daniel, Soon-Jo Chung, and Fred Hadaegh. ``Spacecraft swarm guidance using a sequence of decentralized convex optimizations." (2012): Art-No.
\bibitem{15}
Morgan, Daniel, Soon-Jo Chung, and Fred Y. Hadaegh. ``Swarm assignment and trajectory optimization using variable-swarm, distributed auction assignment and model predictive control." AIAA guidance, navigation, and control conference. 2015.
\bibitem{19}
Mahler, Ronald PS. ``Multitarget Bayes filtering via first-order multitarget moments." IEEE Transactions on Aerospace and Electronic systems 39.4 (2003): 1152-1178.
\bibitem{27}
Vo, B-N., and W-K. Ma. ``The Gaussian mixture probability hypothesis density filter." IEEE Transactions on signal processing 54.11 (2006): 4091-4104.
\bibitem{11}
Todorov, Emanuel, and Weiwei Li. ``A generalized iterative LQG method for locally-optimal feedback control of constrained nonlinear stochastic systems." American Control Conference, 2005. Proceedings of the 2005. IEEE, 2005.
\bibitem{12}
Hoang, Hung Gia, et al. ``The Cauchy Schwarz divergence for Poisson point processes." IEEE Transactions on Information Theory 61.8 (2015): 4475-4485.
\bibitem{13}
Inaba, Masayuki, and Peter I. Corke. ``Linear-quadratic control and smoothing." Robotics Research: The 16th International Symposium ISRR. Switzerland: Springer, 2016. 43-47. Print. 
\bibitem{14}
Findeisen, Rolf, and Frank Allgöwer. ``An introduction to nonlinear model predictive control." 21st Benelux Meeting on Systems and Control. Vol. 11. Eindhoven, The Netherlands: Technische Universiteit Eindhoven Veldhoven, 2002.
\bibitem{30}
Find minimum of unconstrained multivariable function - MATLAB fminunc, MathWorks. Web. 
\end{thebibliography}
\end{document}